\documentclass[11pt, letterpaper]{article}

\usepackage[margin=1in]{geometry} % Standard margins
\usepackage[utf8]{inputenc}
\usepackage[T1]{fontenc}
\usepackage{lmodern} % Improved font rendering
\usepackage{hyperref} % For hyperlinks (emails, references)
\usepackage{graphicx}
\graphicspath{{Fig/}}

\usepackage{appendix}
\usepackage{amsmath}
\usepackage{amssymb}
\usepackage{amsthm}

\usepackage{natbib}

\usepackage{authblk}
\DeclareMathOperator*{\argmin}{arg\,min}

\theoremstyle{plain} % Italic text (Theorems, Propositions)

\theoremstyle{definition} % Upright text (Definitions, Examples)

\theoremstyle{remark} % Upright text (Remarks)

\title{Functional Modeling of Learning and Memory Dynamics in Cognitive Disorders}

\author[1]{Maria Laura Battagliola\thanks{Corresponding author: \texttt{laura.battagliola@itam.mx}}}
\author[2]{Laura J. Benoit}
\author[2]{Sarah Canetta}
\author[3]{Shizhe Zhang}
\author[3]{R. Todd Ogden}

\affil[1]{Department of Statistics, ITAM, Mexico City, Mexico}
\affil[2]{Department of Psychiatry, Columbia University, New York, USA}
\affil[3]{Department of Biostatistics, Columbia University, New York, USA}

\date{} % Leave empty to omit date, or use \today

\begin{document}

\maketitle

\begin{abstract}
Deficits in working memory, which includes both the ability to learn and to retain information short-term, are a hallmark of many cognitive disorders. Our study analyzes data from a neuroscience experiment on animal subjects, where performance on a working memory task was recorded as repeated binary success or failure data. We estimate continuous probability of success curves from this binary data in the context of functional data analysis, which is largely used in biological processes that are intrinsically continuous. We then register these curves to decompose each function into its amplitude, representing overall performance, and its phase, representing the speed of learning or response. Because we are able to separate speed from performance, we can address the crucial question of whether a cognitive disorder impacts not only how well subjects can learn and remember, but also how fast. This allows us to analyze the components jointly to uncover how speed and performance co-vary, and to compare them separately to pinpoint whether group differences stem from a deficit in peak performance or a change in speed.

\vspace{0.5cm}
\noindent\textbf{Keywords:} Functional data analysis, registration, multivariate functional principal component analysis, permutation tests
\end{abstract}

\section{Introduction}
Cognitive impairments are common across many psychiatric disorders, such as schizophrenia. In particular, working memory, namely the capability of assimilating information and storing it for a short period of time in order to perform a task, is a vital component of mental processing. Given how fundamental working memory is to many cognitive deficits, extensive neuroscience research has sought to identify its physical basis in the brain. These efforts have pointed to the connection between the quality of working memory and the physiology of prefrontal regions, particularly the dorsolateral prefrontal cortex in humans and its rodent homolog, the medial prefrontal cortex (mPFC).

Building on this connection, \cite{Benoit2020Medial} investigated the impact of mPFC lesions on working memory during repeated tasks. Specifically, they divided a cohort of mice into two groups: a control group that received a sham surgery, and a second group in which an excitotoxic lesion was induced in the mPFC to disrupt its function. Both groups were trained to perform a delayed non-match to sample task. In this task, mice first pressed one of two levers, then after a variable delay, had to remember this initial choice and press the opposite lever to receive a reward. The initial training, called the acquisition stage, was designed to analyze differences in the learning process between the two groups. Once the mice learned the task, their working memory was challenged by introducing a random delay of 2, 4, 8, or 16 seconds (denoted by s from here on) between the initial choice and the final action. The main research objective was to assess any effect of the mPFC lesion on performance, i.e., whether it impacted either the speed to reach proficiency, or the maximum probability of success achievable, or both.

\begin{figure}[htb!]
\centering
\includegraphics[width=\textwidth]{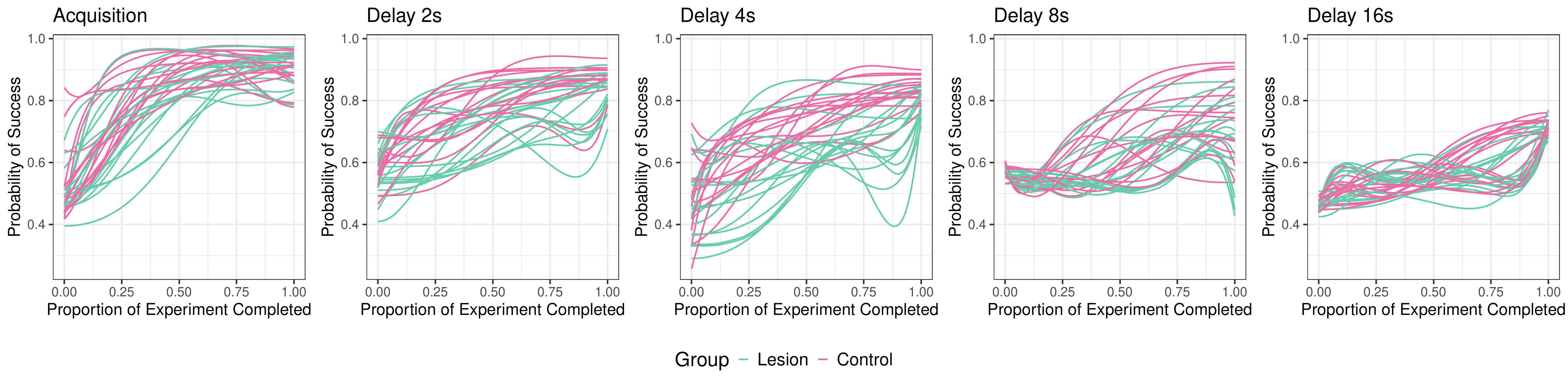}
\caption{Learning curves for all phases of the experiment. The functions belonging to the lesion group are in teal, and those belonging to the control group are in pink.}
\label{fig:curves_intro}
\end{figure}

 The analysis of \cite{Benoit2020Medial} relies on data aggregated by experimental stage, namely the acquisition phase before reaching criterion performance, or the subsequent delay-testing phase, and delay length. While this is appropriate to address their research questions, such an approach does not model the continuous evolution of learning and memory over time. To capture these underlying dynamics, we propose an approach that builds on methods from functional data analysis (FDA), the area of statistics dedicated to observations that can be regarded as curves over some continuum \citep{ramsay2005}. This framework is particularly suitable for biological phenomena like learning, which are intrinsically continuous but are measured at discrete, high-frequency intervals in practice. Moreover, in our case, FDA is also a convenient modeling paradigm, given that recent years have seen the adaptation of a plethora of multivariate methods to functional data. For instance, Principal Component Analysis (PCA) has been generalized to Functional PCA (FPCA) (see, for instance, \cite{yao2005, xiao2013}), and permutational multivariate analysis of variance has been adapted into functional permutation tests \citep{shieh2023permutation}. Thus, we will adapt such FDA-based methods to estimate and then analyze the continuous probability of success shown in Figure \ref{fig:curves_intro}.

Specifically, our approach targets the main question posed by \cite{Benoit2020Medial} by separating each mouse's learning curve into amplitude (overall performance) and phase (speed) components using registration on the binary observations \citep{wrobel2019}. This decomposition allows us to investigate two key aspects of performance. First, we identify ``typical'' performance templates for each experimental stage using bivariate FPCA, by which the data corresponding to each mouse will be represented as a bivariate object including both amplitude and phase information \citep{happ2019}. Second, we formally assess whether the control and lesion groups differ in their overall learning curves and in their separated amplitude and phase components. This is accomplished both globally using permutation tests and locally based on interval-wise testing procedures \citep{pini2017}. In particular, the latter determines not only whether the groups differ in performance and speed, but can also indicate when during the experiment those differences emerge.

Our findings are consistent with the primary conclusions of \cite{Benoit2020Medial}. As an example of this agreement, our analysis also confirms that significant performance differences exist between the lesion and control groups during the acquisition and the memory challenges at the short delays of 2s and 4s. Beyond this confirmation, however, our functional methodology provides a deeper understanding of these effects. A key contribution, for instance, is our ability to decompose this overall difference, revealing whether it is primarily driven by performance level or learning speed, and to identify the intervals where this divergence occurs. Furthermore, as another example, our analysis of the main modes of variation uncovers that rapid learning and high performance are intrinsically linked throughout all experimental stages. These more detailed insights, possible due to the separation of learning speed from performance, would be inaccessible using classical statistical approaches that rely on data aggregation.

This paper is organized as follows. In Section \ref{sec:data} we describe the original binary data and the procedure to estimate continuous learning curves from it. Then, Section \ref{sec:fda_methods} is dedicated to the overview of the methodology used to analyze the functional samples; results are discussed in Section \ref{sec:results}. In Section \ref{sec:discussion} we compare our results to those of \cite{Benoit2020Medial}. Finally, in Section \ref{ref:conclusions} we summarize our analysis and mention possible further developments. Additional results are provided in the appendix.

\section{Data and objectives}
\label{sec:data}
In neuroscience, mouse models are frequently used to understand cognitive processes such as learning and memory. The study introduced by \cite{Benoit2020Medial} investigates how a specific brain region, the medial prefrontal cortex (mPFC), affects performance on a task designed to test these abilities. The experiment's core objective was to determine whether damaging the mPFC in mice would impair their ability to learn and correctly perform the task, thereby validating the task's dependency on this brain region.

\subsection{Original data description}
The experiment involved two distinct groups of mice: a lesion group, which had the mPFC surgically damaged, and a sham-operated control group, which underwent a placebo surgery. The two groups consist of  $n_L = 17$ and $n_C = 16$ mice, respectively.

Data collection was carried out in two stages. The first was an acquisition phase, during which mice learned the task rule over 19 consecutive days of training. During the first 15 days of this phase, 120 trials were recorded per day for each mouse, after which the number was increased to 160 trials per day for the final 4 days. The second stage was a delay phase, during which mice were tested with delays of 2s, 4s, 8s, and 16s introduced into the task to challenge their memory. Within each daily session of this phase, all four delay conditions were randomly interspersed. During this phase, 160 trials were recorded per day.

For each trial, the mouse's choice was recorded as a binary outcome: correct or incorrect. Therefore, the dataset for analysis consists of these binary performance records for each mouse, along with a label indicating which of the two groups it belongs to. Thus, the data available is has the form
\begin{equation*}
    Y_{i,t_{ij}}^{g,d} = \begin{cases}
        1,\; \text{if mouse }i \text{ in group }g \text{ with delay }d \text{ at time }t_{ij} \text{ was correct},\\
        0, \: \text{otherwise},
    \end{cases}
\end{equation*}
where we consider $i=1,\dots,n$ mice, with $n = n_L + n_C$, each belonging to group $g \in \{L, C\}$. Each mouse $i$ was tested at a time $t_{ij}$, allowing for different time records between animals, for different delays $d \in \{0, 2,4, 8,16\}$, where we $d=0$ corresponds to the acquisition phase. Thus, $j=1,\dots, N_{id}$, where $N_{id}$ is the number of trials performed with delay $d$ by mouse $i$.

\subsection{Previous statistical analysis}
For the original analysis, the raw binary data was preprocessed by aggregating the trial outcomes into the percentage of correct trials. This was carried out for each day and for each experimental stages. In particular, during the acquisition, performance was summarized for each mouse across all trials each day. During the delay phase, performance was calculated for each mouse per delay length, averaging across the final four days of testing.
The statistical analysis aimed to determine whether the mPFC lesions impacted the speed and amount of learning. Additionally, a question of interest was whether mice began the acquisition phase at chance-level performance, corresponding to a probability of success of $0.5$, which the day-aggregated analysis could not establish definitively due to within-day learning.

For the acquisition, \cite{Benoit2020Medial} found that mice in the lesion group were significantly slower at learning the task. All mice were trained until they reached a common threshold level of performance, so by the end of acquisition both groups performed equivalently. However, the lesioned mice required significantly more trials to reach this criterion, indicating slower learning. This was quantified using a criterion-based approach, where learning was defined as achieving 3 consecutive days above a threshold of $0.8$ probability of success. While intuitive, this approach reduces the continuous learning process to a single binary milestone. When a delay was introduced, the performance of the two groups tended to improve over time for delays of 2s, 4s, and 8s. In addition, there was a significant difference in the performance of the two groups for shorter delays, whereas the control group performs better than the lesion group. For the 8s delay, there was no significant difference in the performance of the two groups, and for the 16s delay, while there was a significant main effect of time on performance for both groups, there was no significant difference between lesion and control groups.

The study concluded that, while both groups improved in their performance over time for each delay condition, the mice in the lesion group had worse performance than the control group. In particular, \cite{Benoit2020Medial} concluded that the difference in performance between the two groups during acquisition is due to the fact that the lesioned mice learn more slowly than the control group. As for the delay conditions, they concluded that the difference in performance is caused by an inability to retain information by the mice in the lesion group due to memory impairment. Finally, the authors tested whether other factors, such as a decline in motivation in executing the experiment, had an impact on the performance of the two groups, and concluded that motivation and motor capabilities were comparable in the two groups. This suggests that any observed difference is solely due to the memory impairment associated to the lesion in the mPFC, and not to other factors.

\subsection{Functional data modeling approach}
Due to the nature of the data, we adopt a functional data analysis framework. Rather than summarizing results across days, we consider a continuous learning process, and we define the function $\{\mu^{g,d}_i(s), s \in [0,1]\}$ to be the probability of mouse $i$ in group $g$ making a correct decision at time $s$ under delay condition $d$.  Thus our observed data consist of observations
\begin{equation}
\label{eq:binary_p}
    Y_{i}^{g,d}(s) = \begin{cases}
        1,\; \text{if mouse }i \text{ in group }g \text{ with delay }d \text{ at progression }s \text{ was correct},\\
        0, \: \text{otherwise}.
    \end{cases}
\end{equation}
In our implementation we regard the index $s$ not to be chronological time but to indicate the position among the sequence of trials.
Thus we model the data as:
\begin{equation}
\label{eq:bernoulli_hp}
   Y_{i}^{g,d}(s) \sim \text{Bernoulli}(\mu^{g,d}_i(s)), \quad s\in [0,1].
\end{equation}
For the sake of readability, in the following methodology explanations, we omit the superscripts indicating group and delay.

In practice, each mouse's progression is recorded over a discrete regular grid $\{s_{i1},\dots,s_{iN_{id}}\}$, where $N_{id}$ is the total number of trials mouse $i$ undertook with a delay $d$.

\subsubsection{Alignment of learning curves}
\label{sec:alignment}

Within this functional framework, we can separate the variability in the learning curves into two components: amplitude and phase. By accounting for the phase component it is possible to distinguish between two mice whose overall learning process is the same but progresses at different rates (which can change over time).  To account for this we propose to register the learning curves, i.e., to determine subject-specific warping functions $\{\gamma_i(\cdot)\}_{i=1}^n$. Each warping function $\gamma_i: [0,1] \to [0,1]$ is a monotonically increasing bijection of the domain that preserves the boundaries. It transforms a common, shared ``internal'' time axis, $s^{*}$, to the observed, subject-specific time axis, $s$, via the relationship $s^{*} = \gamma_i(s)$. 
The aligned function $f^{*}_i(\cdot)$ corresponds to the observed function $f_i(\cdot)$ evaluated on the internal time $s^{*}$, namely $$f^{*}_i(s) = f_i(\gamma_i(s)), \quad s\in [0,1].$$
This decomposition is particularly relevant for our application, as the amplitude component, represented by $\{f^{*}_i(\cdot)\}_{i=1}^n$, characterizes the actual performance once temporal differences are removed, while the phase component, represented by $\{\gamma_i(\cdot)\}_{i=1}^n$, characterizes the individual relative speed or delays in improving performance.

To perform this decomposition, we apply the registration methodology proposed by \cite{wrobel2019}. This method estimates smooth warping functions directly from binary data without a separate pre-smoothing step, assuming that the binary outcomes \eqref{eq:binary_p} are realizations of a Bernoulli process, as in \eqref{eq:bernoulli_hp}. The methodology operates through an iterative process. A feature of this approach is its use of $B$-spline basis expansions to model smooth functions for both amplitude and phase components.

First, conditional on the current temporal alignment, the method employs Generalized FPCA (GFPCA) to estimate a smooth, subject-specific template function. The smoothness of these amplitude components is governed by $K_a$, the number of basis functions used to represent the underlying probability curves. For binary data, this GFPCA step is handled by a variational EM algorithm, which models the latent probability curves directly from the observed outcomes. 

In the second step, these estimated probability curves serve as targets. The algorithm then updates each subject's warping function, whose flexibility is controlled by  the number of basis functions for the phase components, $K_p$. This update finds the monotonic time transformation that best aligns the raw binary data to the template by maximizing the Bernoulli log-likelihood. The algorithm iterates between updating the templates via GFPCA and re-estimating the warping functions until the alignment converges. 

This process yields the final aligned mean curves $\{\hat{\mu}_i^{*}(\cdot)\}_{i=1}^n$ and the corresponding warping functions $\{\hat{\gamma}_i(\cdot)\}_{i=1}^n$, i.e. the amplitude and phase components, respectively.

\section{Functional data analysis methods}
\label{sec:fda_methods}

In this section we detail the statistical framework of the analysis we conduct on the estimated aligned probability curves, their aligned version, and the corresponding warping functions. 

\subsection{Curve transformations}\label{transformations}
First of all, before carrying out any analysis, we need to apply appropriate transformations to the estimated functions. A direct application of methods like Functional Principal Component Analysis (FPCA), based on the assumption that functions reside in $L^2$ functional space, namely the space of square-integrable functions, is not appropriate due to the complex geometric nature of the space they inhabit.

Let us consider the set of warping functions mapping the observed functional coordinate to the internal one:
\begin{equation}
\label{eq:gamma_space}
    \Gamma([0,1]) = \left \{\gamma:[0,1] \to [0,1] \; \vert \;\gamma(0) = 0, \gamma(1) = 1, \gamma \text{ is a diffeomorphism}  \right \}.
\end{equation}
The set of warping functions $\Gamma([0,1])$ is not a vector space, since it is not closed under standard operations like addition or scalar multiplication, properties that are fundamental prerequisites for most statistical analyses with functional data. To overcome this, we follow the framework proposed by \cite{happ2019} and transform the estimated warping functions from their native space \eqref{eq:gamma_space} to a suitable subspace of $L^2([0,1])$. A critical feature of this transformation is that it must be a bijection, allowing any findings from the analysis to be mapped back to the original space of warping functions for interpretation.

The transformation from $\Gamma([0,1])$ is a two-step process. First, for each $i=1,\dots,n$, we consider the first derivative of the warping function $\{\gamma_i(\cdot)\}_{i=1}^n$, $\{\gamma_i'(\cdot)\}_{i=1}^n$. Intuitively, each $\gamma_i'(\cdot)$ can be understood as the instantaneous speed of the process's ``internal'' functional coordinate relative to observed one. Given $s\in [0,1]$, if $\gamma_i'(s)>1$ indicates that the process is being ``fast-forwarded'' to align a compressed section, while if $\gamma_i'(s)<1$ signifies a ``slow-down'' to align a stretched section of the original curve.
 The space of these derivatives is $B^2([0,1]) = \left \{\gamma' \; | \; \gamma \in \Gamma([0,1]) \right \}.$ 
 
 Now, we apply the centered log-ratio (CLR) transform to these density functions. This mapping is an isometric isomorphism, meaning it preserves the geometric structure during the transformation. For a given $\gamma'(\cdot)$, the CLR transform is defined as:
\begin{equation}
\label{eq:psi_trf}
    \eta(\gamma')(s) = \log(\gamma'(s)) -  \int_{0}^1 \log(\gamma'(u)) du, \quad s\in [0,1].
\end{equation}
Notice that, by centering the logarithm of the derivative of the warping function with respect to its mean, the transform $\eta(\cdot)$ represents acceleration and deceleration of the warping, independent of the overall average speed. 
This transform is such that $\eta: B^2([0,1]) \to S^2([0,1])$, where $S^2([0,1]) = \{g \in L^2([0,1]) \; \vert \; \int_0^1 g(u) du = 0 \}$ is a subspace of $L^2([0,1])$. The constraint on the integral of the functions in $S^2([0,1])$ ensures a unique and identifiable representation for each warping function. In order to recover the warping function $\gamma(\cdot)$ we apply the inverse transform
\begin{equation}
\label{eq:inv_psi}
    \gamma(s) = \frac{\int_0^s e^{\eta(u)} du}{\int_0^1 e^{\eta(u)} du}, \quad s\in[0,1].
\end{equation}

Similarly, a probability curve $\mu: [0,1] \rightarrow [0,1]$ does not exist in a vector space. Therefore, we perform our modeling on the logit transform of $\mu(\cdot)$ instead: 
\begin{align}
\label{eq:logit_trf}
    \nu(s)=\log\left(\frac{\mu(s)}{1-\mu(s)}\right), \quad s\in [0,1].
\end{align}
Since $\nu \in L^2([0,1])$, and the logit transform \eqref{eq:logit_trf} is bijective  with inverse transform
\begin{align}
\label{eq:inv_logit_trf}
\mu(s)= \frac{e^{\nu(s)}}{1+e^{\nu(s)}}, \quad s\in [0,1],
\end{align}
we will model the functions $\{\nu_i(\cdot)\}_{i=1}^n$ and map back to $\{\mu_i(\cdot)\}_{i=1}^n$ for interpretation.

Thus, in order to carry out the following analysis, we use the transformed functions $\{ \widehat{\eta}_i(\cdot)\}_{i=1}^n$ applying transformation \eqref{eq:psi_trf} on the warping functions, and transformed functions $\{ \widehat{\nu}_i(\cdot)\}_{i=1}^n$ applying the logit transform \eqref{eq:logit_trf} on the probability functions.

\subsection{Bivariate Functional Principal Component Analysis}
In what follows, we describe the framework for bivariate FPCA. In particular, this will serve to analyze the joint directions of variation of the amplitude and phase components obtained via registration, similarly to \cite{happ2019}. In our case, the two components are transformed as described in Section \ref{transformations}, and thus we consider bivariate objects $\{\widehat{\nu}^{*}_i(\cdot),\widehat{\eta}_i(\cdot) \}_{i=1}^n$.

For a univariate functional zero-mean stochastic process $W \in L^2([0,1])$, whose covariance function is defined as $\Sigma_{WW}(u, v) = \text{cov}(W(u), W(v))$, with $u,v \in [0,1]$, the eigenfunctions $\{\psi_k(\cdot) \}_{k=1}^\infty$ and corresponding eigenvalues $\lambda_1 \geq \lambda_2 \geq \dots >0$ are found through the eigendecomposition
\begin{align}
\label{eq:eigen_dec}
    \int_0^1 \Sigma_{WW}(u,v) \psi_k(u) du = \lambda_k \psi_k(v).
\end{align}
The eigenfunctions are orthonormal and form a basis for $L^2([0,1])$.

Consider now $\{f_i(\cdot) \}_{i=1}^n$, independent observations of $W(\cdot)$. The Karhunen-Loève (KL) expansion allows us to represent observed functions as
\begin{align*}
     f_i(s) = \sum_{k=1}^{\infty} \xi_{ik} \psi_k(s), \quad i=1,\dots,n,
\end{align*}
where each $\xi_{ik}$ is the corresponding score computed as $\xi_{ik} = \int_{0}^1 f_i(u)\psi_k(u) du$. The scores are uncorrelated, and such that $\mathbb{E}[\xi_{ik}] = 0$ and $\text{Var}(\xi_{ik}) = \lambda_k$. The Percentage of Variance Explained (PVE) by the $k$-th component corresponds to $\left(\lambda_k / \sum_{k=1}^\infty \lambda_k \right)\times100\%$.

The truncated version of the KL representation, namely
\begin{equation}
\label{eq:uniKL}
    f_i(s) \approx \sum_{k=1}^{K} \xi_{ik} \psi_k(s), \quad i=1,\dots,n,
\end{equation}
with a finite number of components $K$, is a low-dimensional approximation of each function.

In practice, the $K$ eigenfunctions and eigenvalues are estimated from the sample covariance matrix. This leads to $\{\widehat{\psi}_k(\cdot)\}_{k=1}^K$ and $\widehat{\lambda}_1 \geq \dots \geq \widehat{\lambda}_K >0 $. Finally, in this finite case, the PVE by the $k$-th component is approximated by  $\left(\widehat{\lambda}_k / \sum_{k=1}^K \widehat{\lambda}_k \right)\times100\%$.

A similar analysis can be carried out on observations of multivariate functional processes. Let us consider another univariate zero-mean processes $Z \in L^2([0,1])$, whose covariance function is $\Sigma_{ZZ}(\cdot, \cdot)$. Now, we define the bivariate functional stochastic process $(W,Z) \in L^2([0,1]) \times L^2([0,1])$. The space is endowed with a weighted inner product defined by a positive scalar $D>0$, namely, for $(W,Z),(W^{'},Z^{'}) \in L^2([0,1]) \times L^2([0,1])$
\begin{align*}
  \langle \langle (W,Z ), (W',Z' )\rangle \rangle_{D} = \int_0^1 W(u)W'(u) du + D \int_0^1 Z(u)Z'(u) du.
\end{align*}
The weight $D$ serves as a measure of relative importance of component $Z(\cdot)$ over $W(\cdot)$ in the joint space.
Moreover, we define cross-covariance functions $\Sigma_{WZ}(u,v) = \text{cov}(W(u), Z(v))$ and $\Sigma_{ZW}(u,v) = \text{cov}(Z(u), W(v))$, with $\Sigma_{WZ}(u,v) = \Sigma_{ZW}(v,u)$. The analogous to the eigenproblem \eqref{eq:eigen_dec} in this case is
\begin{align}
\label{eq:eigen_dec_biv}
\begin{cases}
     \int_0^1 \Sigma_{WW}(u,v) \psi_k(u) du  +  D\int_0^1 \Sigma_{WZ}(u,v) \varphi_k(u) du = \lambda_k \psi_k(v),\\
     D\int_0^1 \Sigma_{ZZ}(u,v) \varphi_k(u) du  +  \int_0^1 \Sigma_{ZW}(u,v) \psi_k(u) du = \lambda_k \varphi_k(v),
\end{cases}
\end{align}
with eigenvalues $\lambda_1 \geq \lambda_2 \geq \dots >0$, and bivariate eigenfunctions $\{(\psi_k (\cdot), \varphi_k(\cdot))\}_{k=1}^\infty$, such that $(\psi_k, \varphi_k ) \in L^2([0,1]) \times L^2([0,1])$ with $k=1, 2,\dots.$ Notice that the bivariate eigenfunctions are orthonormal with respect to the weighted inner product:
$$
\langle \langle (\psi_k, \varphi_k), (\psi_j, \varphi_j) \rangle \rangle_D = \int_0^1 \psi_k(u)\psi_j(u)du + D \int_0^1 \varphi_k(u)\varphi_j(u)du = \begin{cases}
    1, \; \text{ if } k=j;\\
    0, \; \text{ otherwise}.
\end{cases}
$$

When we observe independent observations of $(W(\cdot), Z(\cdot))$, namely the sample $\{ (f_i(\cdot), h_i(\cdot))\}_{i=1}^n$, the truncated KL representation of the bivariate observations is
\begin{align*}
    (f_i(s), g_i(s)) = \sum_{k=1}^K \xi_{ik} (\psi_k (s), \varphi_k(s)),
\end{align*}
where each score is computed as $\xi_{ik} = \int_0^1 f_i(u)\psi_k(u) du + D\int_0^1 h_i(u)\varphi_k(u)du$. This is analogous to \eqref{eq:uniKL} presented previously for the univariate case. 

To perform the multivariate FPCA, we adopt the component-wise method proposed by \cite{happ2018}. This approach constructs the final multivariate principal components from the results of separate univariate FPCAs performed on each component of the data. The method is general, and can also accommodate more complex settings where each component of the multivariate functional object is observed on a different domain. Thus, the estimation in practice is carried out using sample covariance and cross-covariance matrices, leading to estimated eigenvalues and eigenfunctions, $\{(\widehat{\psi}_k (\cdot), \widehat{\varphi}_k(\cdot))\}_{k=1}^K$ and $\widehat{\lambda}_1 \geq \dots \geq \widehat{\lambda}_K \geq 0$, respectively. Analogously to the univariate case, we can write the main modes of variation for any $k=1,\dots,K$ as
\begin{align*}
(f(\cdot), h(\cdot))^{-}_k & = \left(\overline{f}(\cdot), \overline{h}(\cdot)\right)  -2\sqrt{\widehat{\lambda}_k} (\widehat{\psi}_k (\cdot), \widehat{\varphi}_k(\cdot)),\\
(f(\cdot), h(\cdot))^{+}_k & = \left(\overline{f}(\cdot), \overline{h}(\cdot)\right) + 2\sqrt{\widehat{\lambda}_k} (\widehat{\psi}_k (\cdot), \widehat{\varphi}_k(\cdot)).
\end{align*}
where, for $s\in[0,1]$, $\overline{f}(s) = \frac{1}{n}\sum_{i=1}^n f_i(s)$ and $\overline{h}(s) = \frac{1}{n}\sum_{i=1}^n h_i(s)$. Moreover, within this framework we can isolate the modes of variation pertaining to either one of the univariate components as
\begin{align}
\label{eq:bi_modes_minus}
    f_k(\cdot)^{-}  = \overline{f}(\cdot)  -2\sqrt{\widehat{\lambda}_k} \widehat{\psi}_k (\cdot),& \quad h_k(\cdot)^{-}  = \overline{h}(\cdot)  -2\sqrt{\widehat{\lambda}_k}  \widehat{\varphi}_k(\cdot)\\
\label{eq:bi_modes_plus}
 f_k(\cdot)^{+}  = \overline{f}(\cdot)  +2\sqrt{\widehat{\lambda}_k} \widehat{\psi}_k (\cdot),& \quad h_k(\cdot)^{+}  = \overline{h}(\cdot)  +2\sqrt{\widehat{\lambda}_k}  \widehat{\varphi}_k(\cdot).
\end{align}
Notice that, even though it is possible to compute the two sets of modes of variation separately, \eqref{eq:bi_modes_minus} and \eqref{eq:bi_modes_plus} should be interpreted together. Finally, it is possible to determine how much each mode of variation contributes to the total variance of the sample by computing the PVE for each principal component.

\subsection{Permutation testing}
In order to formally establish whether there is a difference between the two experimental groups, $g\in \{L,C \}$, we test the difference between their means. Specifically, the tests are applied independently to three sets of estimated functions, namely the unaligned learning curves, the aligned learning curves, and the transformed warping functions. In order to illustrate the methodology, we consider a generic set of functions $\{f_i(\cdot)\}_{i=1}^n$ belonging to $L^2([0,1])$, where each function is associated with a group label. Thus, we consider the lesion group $\{f^L_i(\cdot)\}_{i=1}^{n_L}$, and the control group $\{f^C_i(\cdot)\}_{i=1}^{n_C}$, with $n = n_L + n_C$. 

\subsubsection{Global test}
The first procedure we employ is a nonparametric permutation testing framework \citep{reiss2010fast,shieh2023permutation}. 
This approach is particularly suitable, as it avoids strong distributional assumptions about the functional data. 

In particular, we test the null hypothesis that the mean functions of the two groups are identical, against the alternative that they are not:
\begin{align*}
    H_0: & \; \mathbb{E}\left[f^L(s)\right] = \mathbb{E}\left[f^C(s)\right] \quad \forall s \in [0,1],\\
    H_1: & \; \mathbb{E}\left[f^L(s)\right] \neq \mathbb{E}\left[f^C(s)\right]\quad \text{for some } s \in [0,1].
\end{align*}
The test statistic is built from the sample mean functions of the two groups, $\overline{{f}^L}(s) = \frac{1}{n_L}\sum_{i=1}^{n_L} f_i^L(s)$ and $\overline{{f}^C}(s) = \frac{1}{n_C}\sum_{i=1}^{n_C} f_i^C(s)$, for $s \in [0,1]$. We quantify the magnitude of the difference between these two mean curves with the test statistic
\begin{equation}
\label{eq:ise_obs}
    T = \int_{0}^{1} \left(\overline{{f}^L}(u) - \overline{{f}^C}(u)\right)^2 du.
\end{equation}

To assess the statistical significance of $T$, we generate an empirical null distribution. This is achieved by pooling all $n$ individual functional curves and repeatedly creating new, random partitions of the data. In each permutation $b \in \{1,\dots,B\}$, $n_L$ curves are randomly reassigned to group $L$ and $n_C$ curves to group $C$. For the permuted dataset at iteration $b$, the same statistic \eqref{eq:ise_obs} is computed, $T_{b}$. This process is repeated a large number of times to create an empirical distribution of the test statistic under the assumption that $H_0$ is true.

Finally, the $p$-value is calculated as the proportion of permuted statistics that are greater than or equal to the statistic of the original functional sample:
\begin{align}
\label{eq:p_val_perm}
    p = \frac{\sum_{b=1}^B \mathbf{1}_{T_b \geq T}}{B},
\end{align}
where $\mathbf{1}_{T_b \geq T}$ is the indicator function. A small $p$-value compared to a significance level $\alpha$ indicates that the observed difference between the group mean functions is unlikely to have occurred by chance, providing evidence to reject the null hypothesis in favor of the alternative.

\subsubsection{Interval-wise test}
\label{sec:iwt}
While the global permutation test described previously can determine whether the mean functions of the two groups differ at all, it can not specify the location of any regions in which the means differ. To identify these sub-intervals of significant difference we employ the non-parametric interval-wise testing (IWT) procedure introduced by \cite{pini2017}.

The core of the IWT method is to perform a separate permutation test, identical to the one described before, for every possible sub-interval $\mathcal{I}=(s_1,s_2) \subseteq [0,1]$. In particular, for each $\mathcal{I}$ we test the local null hypothesis that the mean functions are identical within that specific interval:
\begin{align*}
    H_0^{\mathcal{I}}: & \; \mathbb{E}\left[f^L(s)\right] = \mathbb{E}\left[f^C(s)\right] \quad \forall s \in \mathcal{I},\\
    H_1^{\mathcal{I}}: & \; \mathbb{E}\left[f^L(s)\right] \neq \mathbb{E}\left[f^C(s)\right]\quad \text{for some } s \in \mathcal{I}.
\end{align*}
The test statistic for each interval, $T^{\mathcal{I}}$, is the integrated squared error of the sample means restricted to that interval, analogous to Equation~\eqref{eq:ise_obs}. This yields a $p$-value, $p^{\mathcal{I}}$, for every sub-interval of the domain.

From this collection of interval-specific $p$-values, we construct two functions to summarize the evidence against the null hypothesis across the domain. The first is the unadjusted $p$-value function, $p(s)$, which provides a point-wise assessment of significance. It is defined locally at each point $s$ by identifying the highest $p^{\mathcal{I}}$ from the intervals $\mathcal{I}=(s_1, s_2)$ whose extremes are infinitely close to $s$. Formally, this is the limit superior of $p$-values from intervals fully contained within an $\epsilon$-neighborhood of $s$ as that neighborhood shrinks to zero:
\begin{equation}
    p(s) = \lim_{\epsilon\to 0^+} \left( \sup_{\mathcal{I}} \left\{ p^{\mathcal{I}}: s-\epsilon < s_1 < s_2 < s+\epsilon\right\}\right).
\end{equation}
For a more rigorous control of error across intervals, we also define the adjusted $p$-value function, $\tilde{p}(s)$. This is calculated at each point $s$ by taking the supremum of the $p$-values from all intervals that contain that point:
\begin{equation}
    \tilde{p}(s) = \sup_{\mathcal{I} \ni s} p^{\mathcal{I}}.
\end{equation}
This adjustment accounts for the multiple testing problem inherent in examining all possible intervals and provides control over the interval-wise error rate.

Finally, to select the domains of significant difference, we threshold the adjusted $p$-value function at a pre-determined significance level $\alpha$. The regions $\{s\in[0,1] \; : \; \tilde{p}(s) \leq \alpha \}$ are identified as the sub-intervals where the mean functions of the two groups exhibit a statistically significant difference.

\section{Results}
\label{sec:results}
We now analyze the binary data from \cite{Benoit2020Medial} within the functional framework outlined in the previous sections. 

\subsection{Preprocessing}
We prepared the binary data such that the observation times of the binary outcomes of each trial were chronologically consistent. For each mouse and experimental stage, we normalized the trial observation times to the interval $[0,1]$. This scaling set the first observation at $s=0$ and the last at $s=1$, while preserving the relative time differences between all trials, including those on separate days. 

Then, the registration was carried out with function \texttt{register\_fpca} in the package \texttt{registr} \citep{registr_package}. We used the variational EM algorithm described in Section \ref{sec:alignment} for estimation. In order to isolate the primary learning trends from trial-to-trial noise, we set a low number of $B$-splines for the representation of amplitude and phase components, specifically $K_a = K_p = 4$. Moreover, so as to gain a high degree of information about the phenomena we are studying, we obtaine the first 10 principal components of the GFPCA. 

In order to perform tests later on, we then interpolate each sample of curves on a common dense regular grid on $[0,1]$. The grid size is determined by the minimum number of trials among mice, i.e. $N_d = \min_i N_{id}$ for $d\in \{0, 2,4,8,16 \}$.
The learning (``acquisition'') phase consisted of many more trials than any of the subsequent tests on memory. In particular, for the acquisition phase there were $N_0=2022$ trials. For the tasks with delay we have the following number of trials: $N_2=175$ for $d=2$, $N_4=N_8=170$ for $d=4$ and $d=8$, and $N_{16}=164$ for $d=16$.

\begin{figure}[htb!]
  \centering
\includegraphics[width=\textwidth]{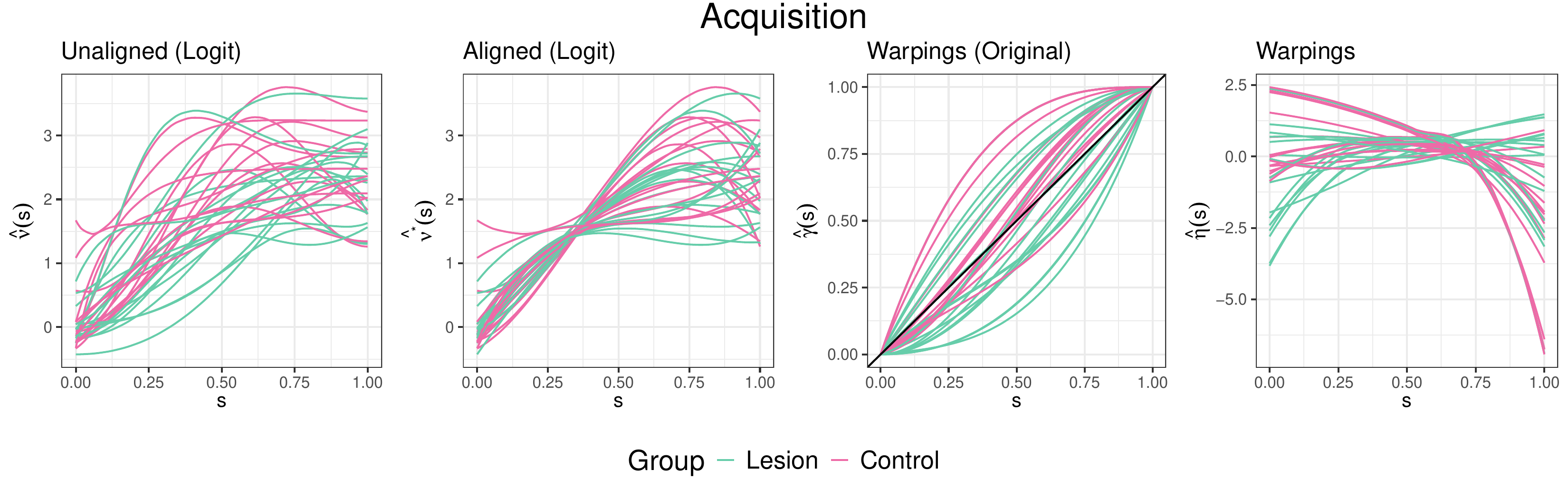}
  \caption{Functional samples for acquisition. We show the logit-transformed learning curves (first panel), their aligned version (second panel), the warping functions of the alignment (third panel), and their transformations (right panel). Lesion mice are colored in teal, and control mice in pink.}
  \label{fig:acq_sample}
\end{figure}

In Figure \ref{fig:acq_sample} we show the functional samples we obtain from preprocessing the results of the registration analysis. Throughout our analysis we refer to the logit transformed learning curves observed over the chronological time, $\{\widehat{\nu}_i(\cdot)\}_{i=1}^n$, as the ``unaligned'' curves, and to their version when observed on the internal time, $\{\widehat{\nu^{*}}_i(\cdot)\}_{i=1}^n$, as the ``aligned'' curves. The ``original warpings'' are the warping functions $\{\widehat{\gamma}_i(\cdot) \}_{i=1}^n$, and we carry out our study on the ``warpings'' $\{\widehat{\eta}_i(\cdot) \}_{i=1}^n$, namely the CLR derivatives of the warping functions described in Section~\ref{transformations}. Notice that the estimated unaligned probability curves confirm that mice begin the acquisition phase at approximately chance-level performance, indicating no prior task knowledge. This insight, also visible in Figure~\ref{fig:curves_intro}, was not accessible from the original day-aggregated analysis due to within-day learning.

The functional samples for the memory tests, into which delays are introduced, are shown in Figure~\ref{fig:delays_sample} in the appendix.

\subsection{Learning Curves Templates}
\label{sec:learning_templates}

In this section, we analyze the main modes of variation obtained with the bivariate FPCA carried out on the curves $\{ (\widehat{\nu^{*}}_i(\cdot),\widehat{\eta}_i(\cdot))\}_{i=1}^n$. For the analysis, carried out with function \texttt{MFPCA} from the \texttt{homonymous} package \citep{MFPCA_package}, we choose $K=10$ components.

We extract the eigenvalues $\widehat{\lambda}_1 \geq \dots \geq \widehat{\lambda}_K \geq 0$, and the bivariate eigenfunctions $\{ (\widehat{\psi}_k(\cdot), \widehat{\phi}_k(\cdot)) \}_{k=1}^K$. In particular, the univariate elements of the eigenfunctions, $\{ \widehat{\psi}_k(\cdot) \}_{k=1}^K$ and $\{ \widehat{\phi}_k(\cdot) \}_{k=1}^K$, correspond to the amplitude and phase components, respectively. We are interested in isolating the amplitude and phase variations using \eqref{eq:bi_modes_minus} and \eqref{eq:bi_modes_plus}. 

To compute the amplitude ($A$) modes of variation, we first define the modes of variation for the aligned logit-curves, $\widehat{\nu}^{*,-}_k(\cdot)$ and $\widehat{\nu}^{*,+}_k(\cdot)$, which represent perturbations around the mean aligned function $\overline{\nu^{*}}(\cdot)$:
\begin{align*}
    \widehat{\nu}^{*,-}_k(s) = \overline{\nu^{*}}(s) - 2\sqrt{\widehat{\lambda}_k} \widehat{\psi}_k(s), \quad \widehat{\nu}^{*,+}_k(s) = \overline{\nu^{*}}(s)+2\sqrt{\widehat{\lambda}_k} \widehat{\psi}_k(s).
\end{align*}
Next, we compose these aligned modes with the mean warping function, $\overline{\gamma}(s)$, and then transform the result into learning curves using the inverse logit transformation \eqref{eq:inv_logit_trf}. The mean warping function is given by $\overline{\gamma}(s) = \frac{\int_0^s e^{\overline{\eta}(u)} du}{\int_0^1 e^{\overline{\eta}(u)} du}$.
This yields the final amplitude modes of variation:
\begin{align*}
   \widehat{\mu}^{-}_{A,k}(s) = \frac{\exp\left(\widehat{\nu}^{*,-}_k \left( \overline{\gamma}(s)\right) \right)}{1 + \exp\left( \widehat{\nu}^{*,-}_k \left( \overline{\gamma}(s)\right) \right)}, \quad \widehat{\mu}^{+}_{A,k}(s) = \frac{\exp\left( \widehat{\nu}^{*,+}_k \left( \overline{\gamma}(s)\right) \right)}{1 + \exp\left( \widehat{\nu}^{*,+}_k \left( \overline{\gamma}(s)\right) \right)}.
\end{align*}
The phase ($P$) modes of variation are computed by composing the mean aligned logit-curve, $\overline{\nu^*}(\cdot)$, with perturbed warping functions. This isolates the effect of temporal shifts in the data. First, we define the perturbed warping functions, $\widehat{\gamma}^{-}_k(\cdot)$ and $\widehat{\gamma}^{+}_k(\cdot)$, by applying the inverse mapping \eqref{eq:inv_psi} to the perturbed mean tangent vector, $\overline{\eta}(\cdot)$, namely
\begin{align*}
   \widehat{\gamma}^{-}_k(s) = \frac{\int_0^s \exp\left(\overline{\eta}(u) - 2\sqrt{\widehat{\lambda}_k}\widehat{\phi}_{k}(u)\right) du}{\int_0^1 \exp\left(\overline{\eta}(u) - 2\sqrt{\widehat{\lambda}_k}\widehat{\phi}_{k}(u)\right) du}, \quad \widehat{\gamma}^{+}_k(s) = \frac{\int_0^s \exp\left(\overline{\eta}(u) + 2\sqrt{\widehat{\lambda}_k}\widehat{\phi}_{k}(u)\right) du}{\int_0^1 \exp\left(\overline{\eta}(u) + 2\sqrt{\widehat{\lambda}_k}\widehat{\phi}_{k}(u)\right) du}.
\end{align*}
Then, the phase modes are then found by composing the mean aligned function $\overline{\nu^{*}}(\cdot)$ with these perturbed warps and applying the transform \eqref{eq:inv_logit_trf}:
\begin{align*}
   \widehat{\mu}^{+}_{P,k}(s) = \frac{\exp \left( \overline{\nu^{*}} \left( \widehat{\gamma}^{+}_k(s) \right) \right)}{1 + \exp \left( \overline{\nu^{*}} \left( \widehat{\gamma}^{+}_k(s) \right) \right)}, \quad 
   \widehat{\mu}^{-}_{P,k}(s) = \frac{\exp \left( \overline{\nu^{*}} \left( \widehat{\gamma}^{-}_k(s) \right) \right)}{1 + \exp \left( \overline{\nu^{*}} \left( \widehat{\gamma}^{-}_k(s) \right) \right)}.
\end{align*}
Moreover, the joint effect of amplitude and phase variation are computed as 

\begin{align*}
   \widehat{\mu}^{+}_{k}(s) = \frac{\exp \left( \widehat{\nu}^{*,+}_k \left( \widehat{\gamma}^{+}_k(s) \right) \right)}{1 + \exp \left( \widehat{\nu}^{*,+}_k \left( \widehat{\gamma}^{+}_k(s) \right) \right)}, \quad 
   \widehat{\mu}^{-}_{k}(s) = \frac{\exp \left( \widehat{\nu}^{*,-}_k \left( \widehat{\gamma}^{-}_k(s) \right) \right)}{1 + \exp \left( \widehat{\nu}^{*,-}_k \left( \widehat{\gamma}^{-}_k(s) \right) \right)},
\end{align*}
accounting for the composition of the two effects. Finally, the overall mean function is computed as
\begin{align*}
    \overline{\widehat{\mu}}(s) = \frac{\exp \left( \overline{\nu^{*}} \left( \overline{\gamma}(s) \right) \right)}{1 + \exp \left( \overline{\nu^{*}} \left( \overline{\gamma}(s)  \right) \right)}. 
\end{align*}

Notice that, in order to compute the marginal and joint modes of variation, one needs to establish the weighting constant $D$ in the eigenproblem \eqref{eq:eigen_dec_biv}. Following the proposed method by \cite{happ2019}, we select $\widehat{D}$ by performing a grid search over a predefined range of candidate values. The optimal weight is the one that minimizes the mean integrated square difference between the unaligned curves and their reconstructed version. Specifically, for a fixed $D$, we can reconstruct the transformed amplitude and phase components for each observation $i=1,\dots,n$ using their respective truncated KL representation, i.e.
\begin{align*}
    \widehat{\nu}^{*}_i(s) &= \overline{\nu^{*}}(s) + \sum_{k=1}^K \widehat{\xi}_{ik} \widehat{\psi}_k(s), \\
    \widehat{\eta}_i(s) &= \overline{\eta}(s) + \sum_{k=1}^K \widehat{\xi}_{ik} \widehat{\varphi}_k(s),
\end{align*}
with $\widehat{\xi}_{ik} = \int_0^1 \left( \widehat{\nu}^{*}_i(u) - \overline{\nu^{*}}(u) \right) \widehat{\psi}_k(u) du + D \int_0^1 \left( \widehat{\eta}_i(u) - \overline{\eta}(u) \right) \widehat{\varphi}_k(u) ds
$. By transforming the phase components to warping functions $\widehat{\gamma}_i(s) = \frac{\int_0^s \exp(\widehat{\eta}_i(u)) \,du}{\int_0^1 \exp(\widehat{\eta}_i(u)) \,du}$, we obtain approximate learning curves
$$
\widehat{\mu}^{(D)}_i(s) = \frac{\exp\left( \widehat{\nu}^{*}_i\left(\widehat{\gamma}_i(s)\right) \right)}{1 + \exp\left( \widehat{\nu}^{*}_i\left(\widehat{\gamma}_i(s)\right) \right)},
$$
where we made the dependence on $D$ explicit. Thus, we compute the Mean Integrated Square Error (MISE) as
$$
\text{MISE}(D) = \frac{1}{n} \sum_{i=1}^n \int_0^1 \left( \mu_i(s) - \widehat{\mu}^{(D)}_i(s) \right)^2 ds.
$$
The optimal weight minimizes the MISE, i.e.
$$\widehat{D} = \argmin_{D \in \mathcal{G}}\ \text{MISE}(D),$$
where the minimization is performed over the discrete grid of candidate values $\mathcal{G} = \{0.1,0.2,\dots,5.0 \}$. Thus, the optimal value $\widehat{D}$ provides an insight into the performance of the mice, quantifying the relative importance of the amount of success (amplitude) versus the speed to success (phase), and revealing which factor is the more dominant source of variation in the experiment. 

In Table \ref{tab:C.hat} we report the estimated values. The relative weight of the phase component varies depending on the stage of the experiment. In particular, we notice that in the acquisition $\widehat{D}$ takes the maximal value of $\mathcal{G}$, indicating that the speed of learning plays a role five times more important than the actual probability of success achieved in the performance of mice. On the contrary, when the shortest and the longest delays, 2s and 16s respectively, are introduced, the estimated weighting constant is minimal, signaling that for these delays the performance of mice is mainly determined by the amplitude component. For delays of 4s and 8s, $\widehat{D}$ takes values around the middle of the grid, showing that the phase component is more predominant.

\begin{table}[htb!]
\centering
\caption{Estimated weighting constant $\widehat{D}$ for each stage of the experiment.}
\label{tab:C.hat}
\begin{tabular}{|l|c|c|c|c|c|}
\hline
 & \textbf{Acquisition} & \textbf{Delay 2s} & \textbf{Delay 4s} & \textbf{Delay 8s} & \textbf{Delay 16s} \\
\hline\hline
$\widehat{D}$ & 5.0  & 0.1 & 2.5 & 2.9 & 0.1 \\
\hline
\end{tabular}
\end{table}

\begin{figure}[tb]
  \centering
\includegraphics[width=0.8\textwidth]{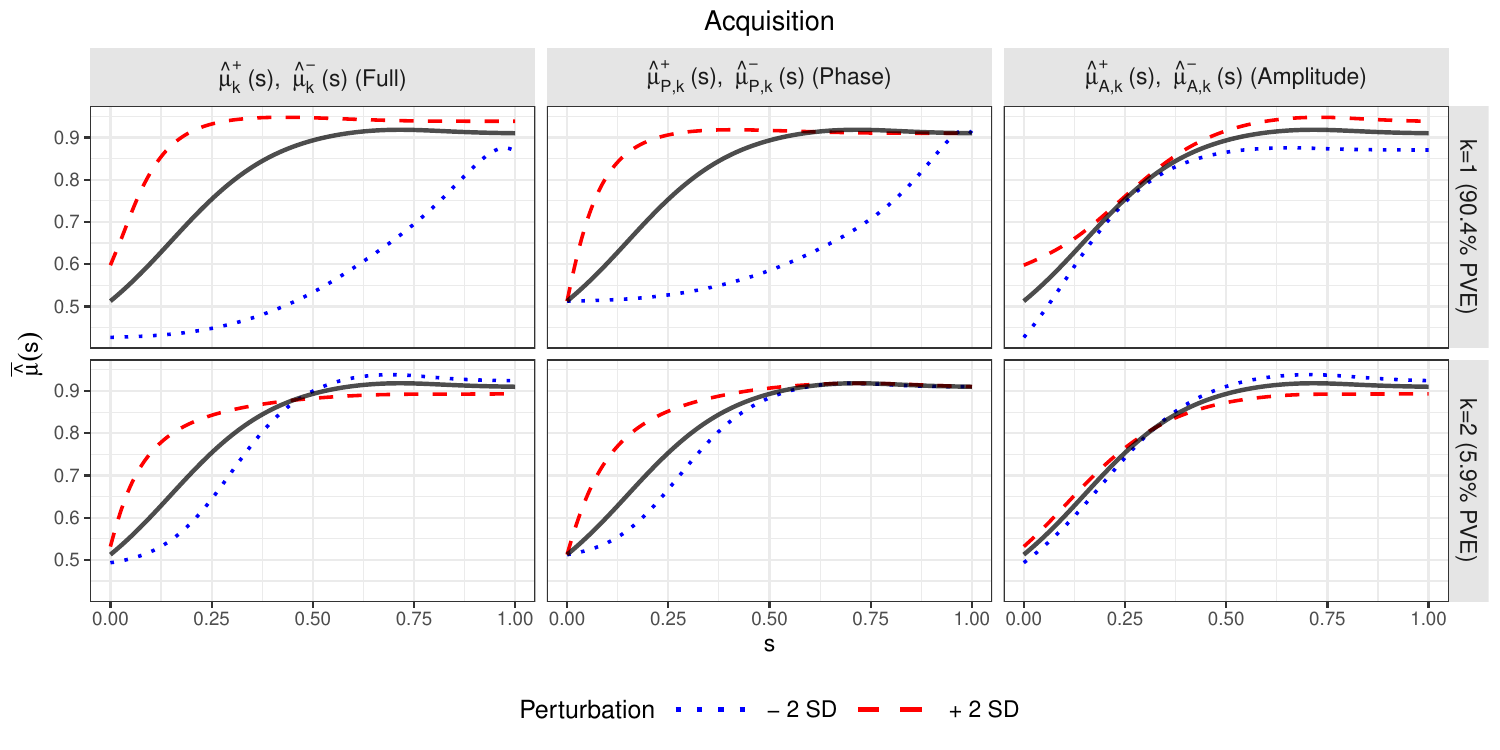}
\caption{Modes of variation around the mean (solid black) given by the first (first row) and second (second row) by the bivariate FPCA analysis for the acquisition. The joint variation (first column), the phase variation (second column) and the amplitude variation (third column) are displayed. The percentage of variance explained (PVE) by each component is shown as well.}
  \label{fig:modes_var_learning}
\end{figure}

After adjusting for the relative weight of phase on amplitude by $\widehat{D}$, we proceed to compute the modes of variation around the mean determined by the first two bivariate eigenfunctions. In Figure \ref{fig:modes_var_learning} we show the modes of variation for acquisition. We notice that the first component, which explains more than 90\% of the total variance of the sample, represents profiles that learn faster than the average mouse, plateauing already at the end of the first quarter of the experiment, while achieving better results than the average, especially at the beginning and throughout the second half of the experiment. This translates to the joint, or full, modes of variation, where the template learning curve is that of a better than average performance, both in terms of probability of success achieved and speed of learning. The second component is characterized by a learning speed profile that is higher than average only in the first half of the experiment, associated with a probability of success that is slightly higher and slightly lower than average in the first and second parts of the acquisition, respectively. The corresponding full mode of variation is that of a mouse that learns faster and has better chances of achieving higher results than the average in the first half of the process of learning, but then does not improve in the second half.

\begin{figure}[tb]
  \centering
\includegraphics[width=0.6\textwidth]{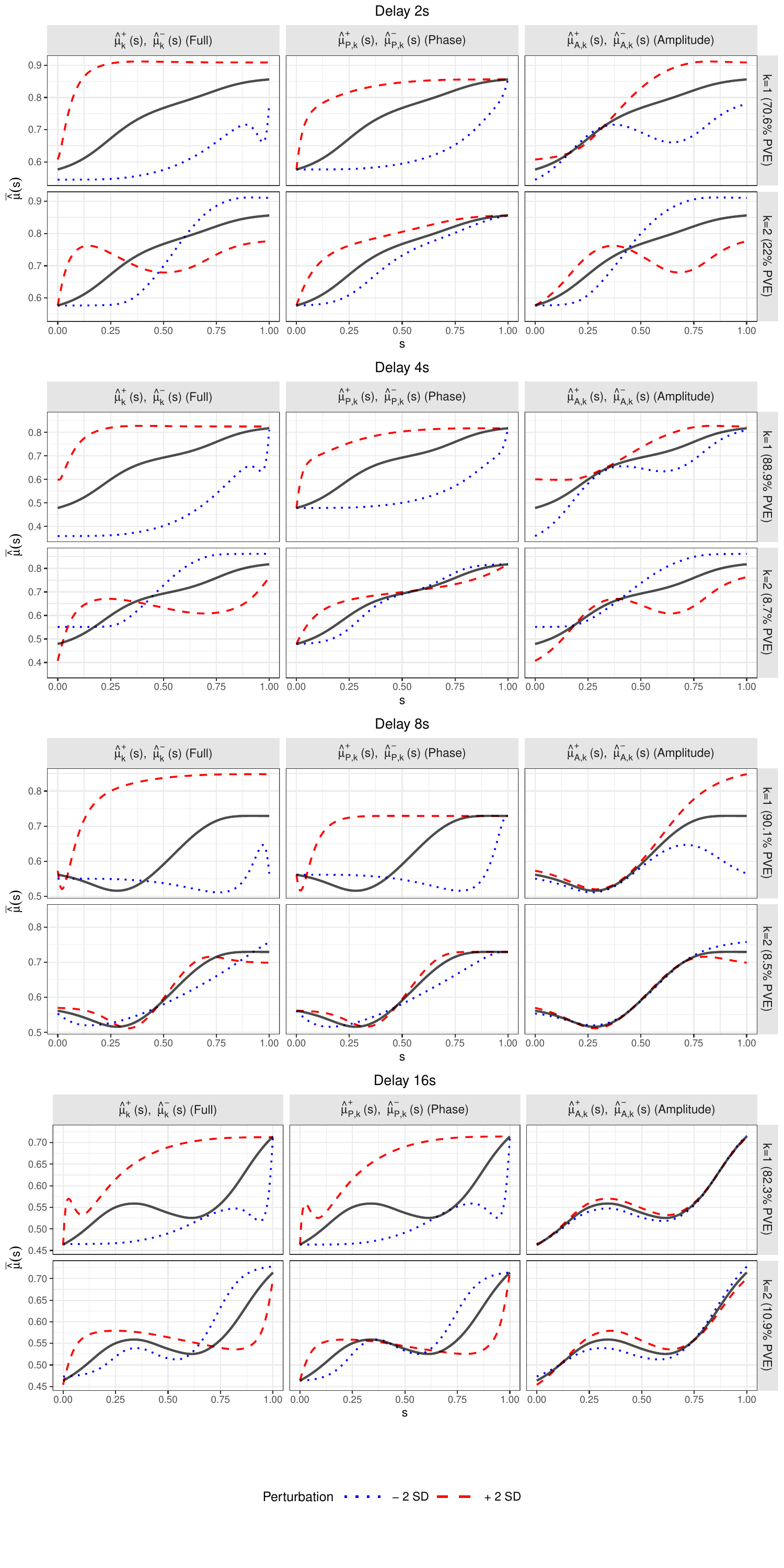}
  \caption{Modes of variation around the mean (solid black) given by the first (first row) and second (second row) by the bivariate FPCA analysis for each delay. The joint variation (first column), the phase variation (second column) and the amplitude variation (third column) are displayed. The percentage of variance explained (PVE) by each component is shown as well.}
  \label{fig:modes_var_memory}
\end{figure}

Figure \ref{fig:modes_var_memory} shows the modes of variation corresponding to the first two bivariate FPCA components for each delay. First, we notice that the first mode of variation, phase and amplitude, and thus joint, of the shortest delays, i.e. delays 2s and 4s, are similar to that acquisition. In particular, it shows the performance of a mouse that reaches a higher-than-average plateau fast. In particular, the amplitude components show that there is more variability at the beginning and in the second half of the experiment,  just as for the acquisition. The first phase mode of variation for the longer delays of 8s and 16s are also similar to the previous ones, showing a performance template that is faster than the average. However, the amplitude components highlight variation in different regions of the domain, namely in the second half for delay 8s, and in $s \in [0.25, 0.75]$ for delay 16s.

The second components correspond to two distinct patterns. The first involves a learning speed that is initially faster than average before slowing down, paired with high success rates in specific areas like the 4s and 16s delays. The second pattern shows a consistently high learning speed, but with performance that is strong at the beginning and then declines, as seen with the 2s delay. The modes of variation associated to the 8s delay are less clearly interpretable, showing a more erratic pattern around the average performance. They result in a joint mode of variation that is either below or above average depending on the portion of domain considered.

\clearpage

In conclusion, after accounting for the relative weights of phase and amplitude being different, the main patterns of variation are remarkably similar across all experimental stages, consistently identifying a profile that reaches high success rates more quickly than average. This finding highlights a connection between the speed of learning and the peak performance achieved, suggesting they are intertwined aspects of performance. This pattern is underscored by the first bivariate component, which explains the vast majority of the total variance across all datasets analyzed. However, we note that the apparent deceleration in learning curves as performance approaches high levels is partially a consequence of our modeling approach. The inverse logit transformation \eqref{eq:inv_logit_trf} used to recover probability curves naturally compresses differences near the boundaries, and thus the flattening observed at high performance levels reflects both genuine ceiling constraints on task performance and properties of the back-transformation.

Furthermore, our analysis reveals two additional key trends. First, there is an inverse relationship between the length of the delay and the maximum chance of success. Second, the learning trajectory when delays are introduced, which begins at around 0.5 success rate and increases non-monotonically, differs significantly from the acquisition phase, where performance instead plateaus at a 0.9 success rate midway through the process. Notice that one factor that may contribute to the non-monotonic patterns observed in delay trial performance is trial history. In the original experiment, the order of delay lengths was randomized across trials within each session. If a mouse performs a particularly challenging long-delay trial, this could create confusion that carries over to subsequent trials, regardless of their delay length. This effect may be amplified when the inter-trial interval is short relative to the delay, potentially disrupting the animal's sense of trial structure.

\subsection{Assessing differences between  groups}

The previous analysis identified the overall patterns of the learning curves at each experimental stage. We now compare the performance of the control and lesion groups. This section illustrates the differences between their mean functions, both globally and across specific subintervals of the experimental progression.

\subsubsection{Global tests}
First, we test whether the mean functions of the two groups differ significantly in a global sense. We run the permutation tests on three sets of curves: the set of unaligned curves $\{ \widehat{\nu}_i(\cdot)\}_{i=1}^n$, their aligned version $\{ \widehat{\nu^*}_i(\cdot)\}_{i=1}^n$, and the transformed warping function $\{ \widehat{\eta}_i(\cdot)\}_{i=1}^n$. We ran $B=1000$ permutations, and in Table \ref{tab:perm_tests} we show the values of the $p$-value \eqref{eq:p_val_perm} across delay conditions. 

During the acquisition phase, the overall difference between the groups' unaligned learning curves is close to significance. Decomposing these curves into their amplitude and phase components reveals that this trend is not driven by the peak success probability achieved, as the aligned curves show no significant difference, but rather by the speed of learning, captured by the transformed warping functions. Therefore, the primary distinction between the groups during this learning stage does not lie in their success potential, but rather in the timing to achieve it. This insight would have remained hidden without a registration analysis.

Introducing a memory challenge via delays revealed that the performance difference between the groups is dependent on the delay's length. Significant distinctions emerged only at the shortest delays. For the 2s delay, the groups' overall learning curves were significantly different, an effect driven by a combination of both a lower achievable success rate (amplitude components) and a slower learning speed (phase components). This compounding deficit, namely being systematically worse and slower, resulted in a highly significant overall difference. At the 4s delay, the overall curves remained significantly different, but this distinction was now primarily driven by the pace of learning rather than the peak performance achieved. This suggests that while both groups struggled more with the increased delay, one group reached its performance ceiling much faster than the other. Conversely, for longer delays, 8s and 16s, all significant differences vanished. This shows that the task became too challenging even for healthy mice, who are unable perform better than the lesion group.

\begin{table}[htb!]
\centering
\caption{Comparison of the $p$-values corresponding to the unaligned curves, their aligned counterparts, and transformed warping functions across stages of the experiment.}
\label{tab:perm_tests}
\begin{tabular}{|l|c|c|c|c|c|}
\hline
 & \textbf{Acquisition} & \textbf{Delay 2s} & \textbf{Delay 4s} & \textbf{Delay 8s} & \textbf{Delay 16s} \\
\hline\hline
\textbf{Unaligned Curves} & 0.056 & \textbf{0.005} & \textbf{0.000} & 0.080 & 0.112 \\
\hline
\textbf{Aligned Curves} & 0.350 & \textbf{0.035} & 0.066 & 0.060 & 0.362 \\
\hline
\textbf{Warping functions} & \textbf{0.019} & \textbf{0.034} & \textbf{0.006} & 0.224 & 0.198 \\
\hline
\end{tabular}
\end{table}

In conclusion, while the control group is faster at learning, there is a significant performance gap between the two groups at short delays, stemming from a combined deficit of speed and accuracy (2s) or a primary timing impairment (4s). When the mice must retain information for longer, these differences vanish as the cognitive load becomes equally challenging for both groups, leveling their performance.

These findings invite two possible mechanistic interpretations. The first is that the similar $p$-value function shapes across acquisition and short delays reflect a shared cognitive mechanism for learning and short-term recall, one that breaks down under high cognitive load at longer delays. Under this view, the mPFC supports both processes, and lesioned animals are consistently impaired in both speed and, to varying degrees, peak performance.

An alternative interpretation centers on the possibility that mice recruit different neural pathways depending on task difficulty. The mPFC is specifically implicated in short-term memory maintenance. At the 2s delay, this interval may be short enough that mice continue relying on mPFC-dependent strategies, explaining why lesioned animals exhibit deficits in both learning speed and peak performance. However, at the 4s delay, the memory demand may exceed the capacity of mPFC-dependent mechanisms, prompting all animals to recruit alternative neural pathways. If these alternative circuits are intact in both groups, this would explain why amplitude differences diminish: both groups can ultimately achieve comparable performance using the same non-mPFC strategy. The persistent phase difference would then reflect a general learning disadvantage in lesioned animals rather than a specific memory impairment. Under this interpretation, the similarity in $p$-value function shapes between acquisition and the 4s delay arises because animals are effectively acquiring a new task strategy, rather than because the same cognitive mechanisms are engaged. At longer delays of 8s and 16s, the task may simply exceed the capacity of any available strategy, eliminating detectable group differences.

Distinguishing between these interpretations would require targeted neurobiological investigation, but both accounts are consistent with the statistical patterns observed in our analysis.

\subsubsection{Interval-wise tests on mean functions differences}

Now that we have established that there is a significant difference between the mean functions of the two groups in certain experimental stages, we want to investigate whether there are sub-intervals over which these curves are significantly different. In order to carry out the analysis in practice, each individual functional observation in both groups is represented using cubic $B$-splines. This finite representation allows to carry out the permutation tests over the labels of the splines coefficients. Moreover, the knots of the basis functions coincide with the dense grid over [0,1], allowing to test every possible discretized compact interval over the domain. 
The implementation was carried out with \texttt{R} package \texttt{fdatest} \citep{fdatest_package}.

\begin{figure}[htb!]
  \centering
\includegraphics[width=0.6\textwidth]{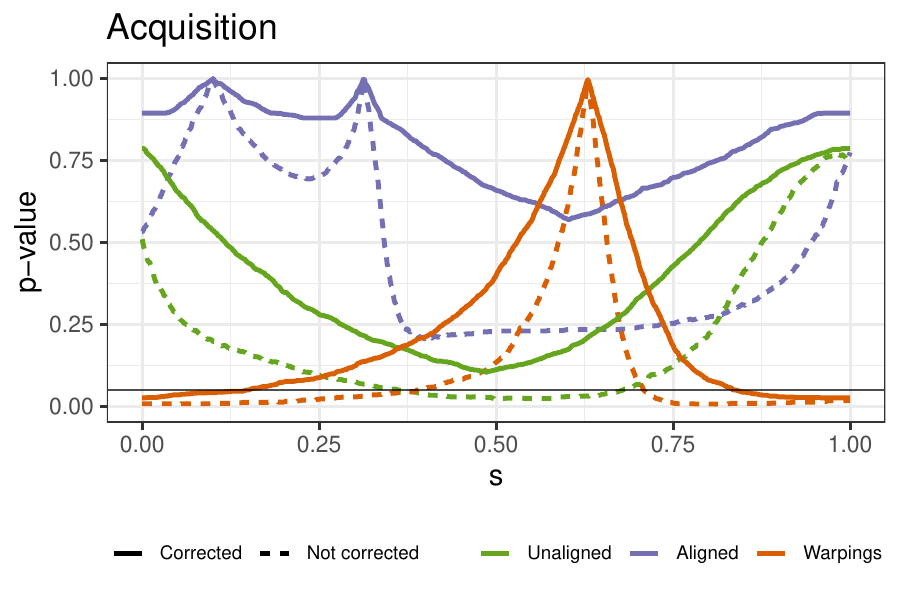}
  \caption{Corrected (solid) and not corrected (dashed) $p$-value functions for unaligned (green), aligned (blue) and transformed warping functions (orange), for the acquisition. The horizontal line corresponds to 0.05.}
  \label{fig:IWT_pvalue_acquisition}
\end{figure}

Figure \ref{fig:IWT_pvalue_acquisition} shows the corrected $p$-value function, $\tilde{p}(\cdot)$, alongside the uncorrected version, $p(\cdot)$, for the acquisition. The unaligned curves, which approached global significance in our previous analysis, mainly differ midway through this experimental stage. This indicates that while both groups start and end at similar performance levels, their learning trajectories differ significantly during the middle of the experiment. In contrast, the transformed warping functions are significantly different at the very beginning and end. This result is a direct consequence of how learning speed is encoded in the phase components. As discussed in Section \ref{sec:learning_templates}, the dominant modes of variation distinguish learners who are consistently faster or slower than the average. A ``faster" learner profile, will exhibit its greatest deviation in speed at the beginning, as it accelerates away from the mean, and near the conclusion, as it finishes early. Therefore, the statistical difference between the groups' learning speeds is most pronounced close to the boundary of the domain, which aligns with the global permutation test results in Table \ref{tab:perm_tests}.

\begin{figure}[htb!]
  \centering
\includegraphics[width=0.8\textwidth]{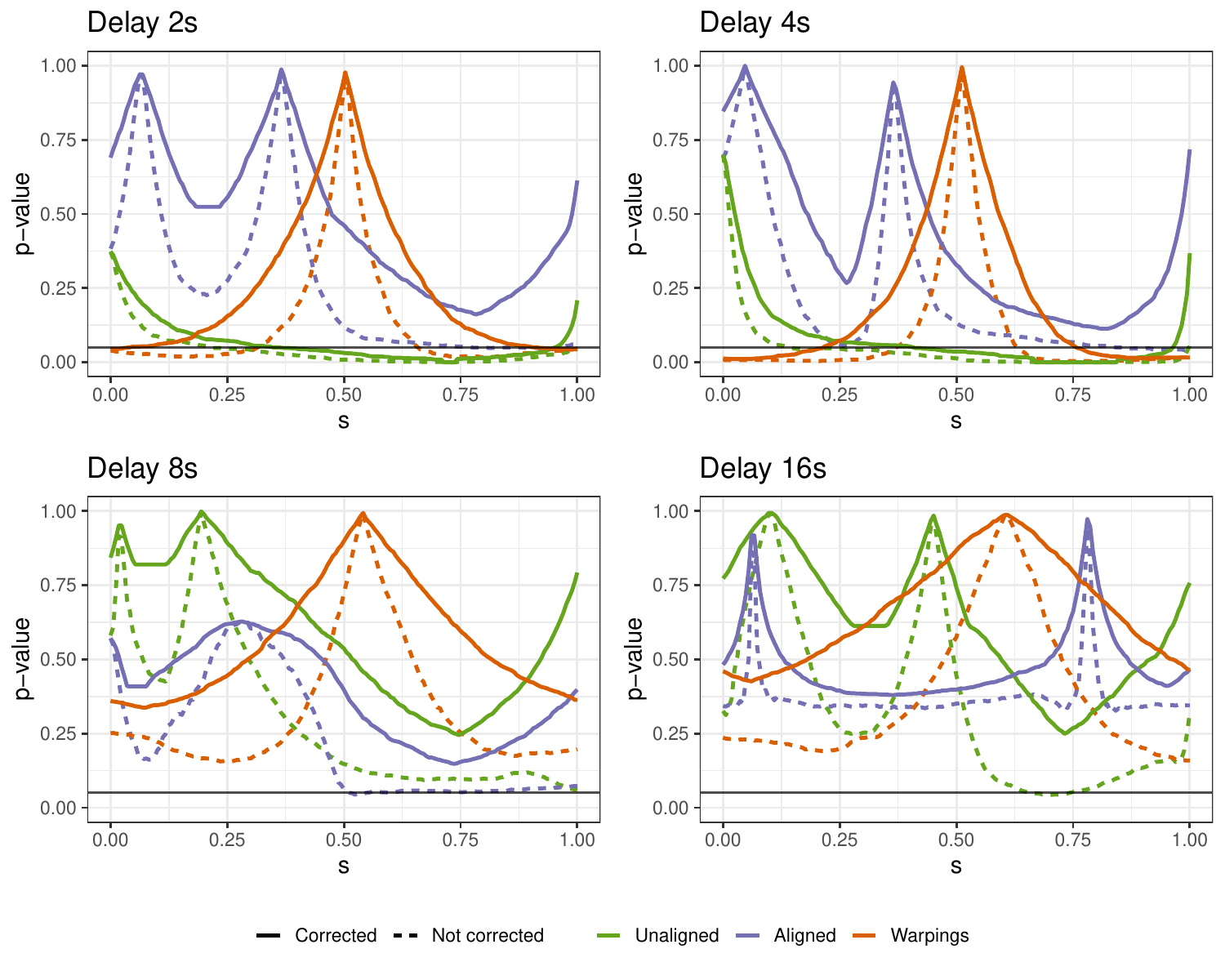}
  \caption{Corrected (solid) and not corrected (dashed) $p$-value functions for unaligned (green), aligned (blue) and transformed warping functions (orange), for the acquisition. The horizontal line corresponds to 0.05.}
  \label{fig:IWT_pvalue_by_delay}
\end{figure}

Figure \ref{fig:IWT_pvalue_by_delay} shows the interval-wise testing results for the memory test stages. For short delays (2s and 4s), the results show similarities with the acquisition phase, as the $p$-value functions exhibit comparable shapes. This suggests that the underlying cognitive mechanisms for initial learning and short-term recall are closely related.

However, a critical shift occurs in the unaligned curves, which are now significantly different primarily in the second half of the experiment. This indicates that the lesion group's performance degrades progressively under a sustained memory load, before they manage to recover slightly at the very end. In contrast, the warping functions remain different at the beginning and end of the stage, reinforcing our earlier conclusion that the main difference in learning speed is concretized in the acceleration and deceleration at close to the boundary of the functional domain.

A different pattern emerges for the longer delays (8s and 16s). While the results remain non-significant, the shape of the $p$-value function for the phase component still resembles that of the earlier stages. Nevertheless, the functions for the unaligned and aligned curves change completely. This divergence suggests that, when the memory load becomes too great, the memorizing strategy changes, and it is replaced by an entirely different cognitive approach.

Comparing Figures \ref{fig:IWT_pvalue_acquisition} and \ref{fig:IWT_pvalue_by_delay}, we notice that the patterns of the $p$-value functions, both corrected and uncorrected, are similar for acquisition and the shorter delays. This shows that the two groups differ in learning and memorizing tasks for short period of times in a similar way. For longer delays, the $p$-value functions corresponding to the original curves and the aligned curves change completely from the previous ones, suggesting a different behavior from the two groups. Finally, we notice that the shape of the $p$-value functions regarding the transformed warping curves always peaks near the middle  of the experiments. We recall that the transformed phase representative \eqref{eq:psi_trf} characterizes the acceleration at which each mouse learn compared to the rest of the sample. Thus, we conclude that the most prominent differences in the learning speed of the two groups are at the beginning and at the end of each experiment.

\section{Discussion}
\label{sec:discussion}

The analysis presented in this work corroborates the primary conclusions of the original study by \cite{Benoit2020Medial}. In particular, we confirmed their finding of an overall performance deficit in mice with mPFC lesions across both learning and memory tasks. 

To address their central question regarding deficits in peak success versus learning speed, our first contribution was to estimate a continuous learning curve for each mouse from the raw binary data. We then registered these functions to decompose into two distinct components: amplitude, which quantifies the peak proficiency achieved, and phase, which captures the speed and timing of learning. We built onto these estimated curves to quantify the exact nature of the impairment. This framework allowed to access deeper insights into the experimental results. For instance, our trial-level approach confirmed that mice begin at chance-level performance, an insight not accessible from day-aggregated analyses.

In agreement with the results of \cite{Benoit2020Medial}, our study shows a close to significant performance gap between the lesion and control groups during the acquisition. When the memory of mice is tested and delays are introduced, we concluded that there is a global significant difference in the performance of the two groups at shorter delays of 2s and 4s. Furthermore, we also support their observation that this difference vanishes at longer delays of 8s and 16s, likely because the task of retaining information for so long is too difficult even for healthy mice.

Our tests on unaligned curves, as well as their phase and amplitude components, revealed the precise nature of these deficits, providing insights beyond the original study. During acquisition, the performance gap was driven almost exclusively by phase, not amplitude, indicating the lesion impairs learning speed rather than ultimate learning capacity. This dynamic shifted during memory tests. The delay of 2s induced a deficit in both amplitude and phase, impacting peak success as well as recall speed. However, at a 4s delay, the impairment was driven only by a difference in phase components. This does not imply equivalent peak performance, but rather that speed differences were more prominent. An alternative interpretation is that at this delay, all mice may recruit non-mPFC pathways, reducing amplitude differences while preserving phase differences due to general learning disadvantages in lesioned animals. The interval-wise tests identified these differences along the development of the experiment: for the 2s and 4s delays, the unaligned curves differed significantly primarily in the second half of the experiment, suggesting progressive performance degradation under memory load. Moreover, the $p$-value functions for acquisition and these short delays shared a similar shape. This pattern disappeared for longer delays, implying a common mechanism for learning and short-term recall that breaks down under high cognitive load. Notice that the warping functions consistently showed significant differences at the beginning and end of the experiment, demonstrating that the groups differ most in their initial learning acceleration and final deceleration.

This consistent $p$-value shape for the transformed warping functions is directly supported by our bivariate FPCA findings. The analysis revealed that the dominant mode of variation, which explains the vast majority of variance across all experimental stages, corresponds to a profile that is both faster and achieves a higher success rate than the average. A mouse fitting this ``faster'' profile naturally exhibits its greatest deviation in learning speed at the start as it accelerates and at the finish as it decelerates. This dominant component confirms that, across all experimental stages, performing better and faster are intrinsically connected. We note that some flattening near high performance levels reflects ceiling constraints as well as properties of the logit back-transformation used in our modeling.

Finally, our estimation of the relative weight between phase and amplitude allowed us to quantify just how prominent the effect of learning speed is on overall performance for each specific experimental stage. Specifically, we found that during acquisition, learning speed was the dominant factor, being five times more important than peak success. Conversely, for the shortest and longest delays, 2s and 16s respectively, this relationship inverted, and performance variation was almost entirely driven by the peak success achieved. For the other delay lengths, learning speed was again more predominant, showing how the nature of the cognitive challenge fundamentally changed with each task. The non-monotonic patterns observed across delay trials may also be influenced by trial history effects, as the randomized order of delay lengths could cause carry-over confusion between trials.

\section{Conclusion}
\label{ref:conclusions}

This work applied a functional data analysis (FDA) framework to investigate the learning and memory dynamics in mice with and without mPFC lesions, using data from \cite{Benoit2020Medial}. We modeled the binary success/failure records from acquisition and memory tasks as continuous probability-of-success curves. The central contribution was the use of curve registration to decompose these trajectories into amplitude and phase components, representing peak proficiency and performance speed, respectively. This separation provided a basis to address the original study's key question: whether cognitive deficits were driven by a reduced performance ceiling or a slower rate of learning. To do so, we employed the registration method from \cite{wrobel2019}, which allows for the direct analysis of binary data without an intermediate smoothing step, and adopted the warping function transformation recommended by \cite{happ2019}. 

While the study presented here is based on these modeling choices, further analysis can be pursued within this functional framework. For instance, it is known that phase-amplitude separation can present identifiability challenges \citep{kneip2008,happ2019, Chakraborty2021}. A complementary future study could therefore explore this dataset using alternative registration algorithms, or different warping function transformations. Such an investigation would provide a valuable comparative perspective. Moreover, the scope of our analysis was primarily constrained by the available sample size. While our approach successfully characterized the key dynamics within this cohort, a larger dataset would provide the necessary statistical power to allow for more integrated modeling. For instance, the fitting of a function-on-scalar regression model, where the success curves could be formally regressed on experimental factors like group and experimental stage. Such model would allow for a further understanding of how these variables directly shape the entirety of the learning and memory trajectories, building directly upon the foundational insights presented in this work. Finally, extending our functional approach to account for trial history effects, as discussed previously, represents another promising direction for future research.

\section{Data availability}
Access to the original data supporting the findings of this study is available upon request.

\section{Competing interests}
The authors declare they do not have competing interests.

\bibliographystyle{apalike}
\bibliography{reference} 

\begin{appendices}
\section{Additional results}

\subsection{Functional samples}
In the following figures we show the functional samples for the memory tests at each delay. These are analogous to Figure \ref{fig:acq_sample}.

\begin{figure}[htb!]
  \centering
\includegraphics[width=\textwidth]{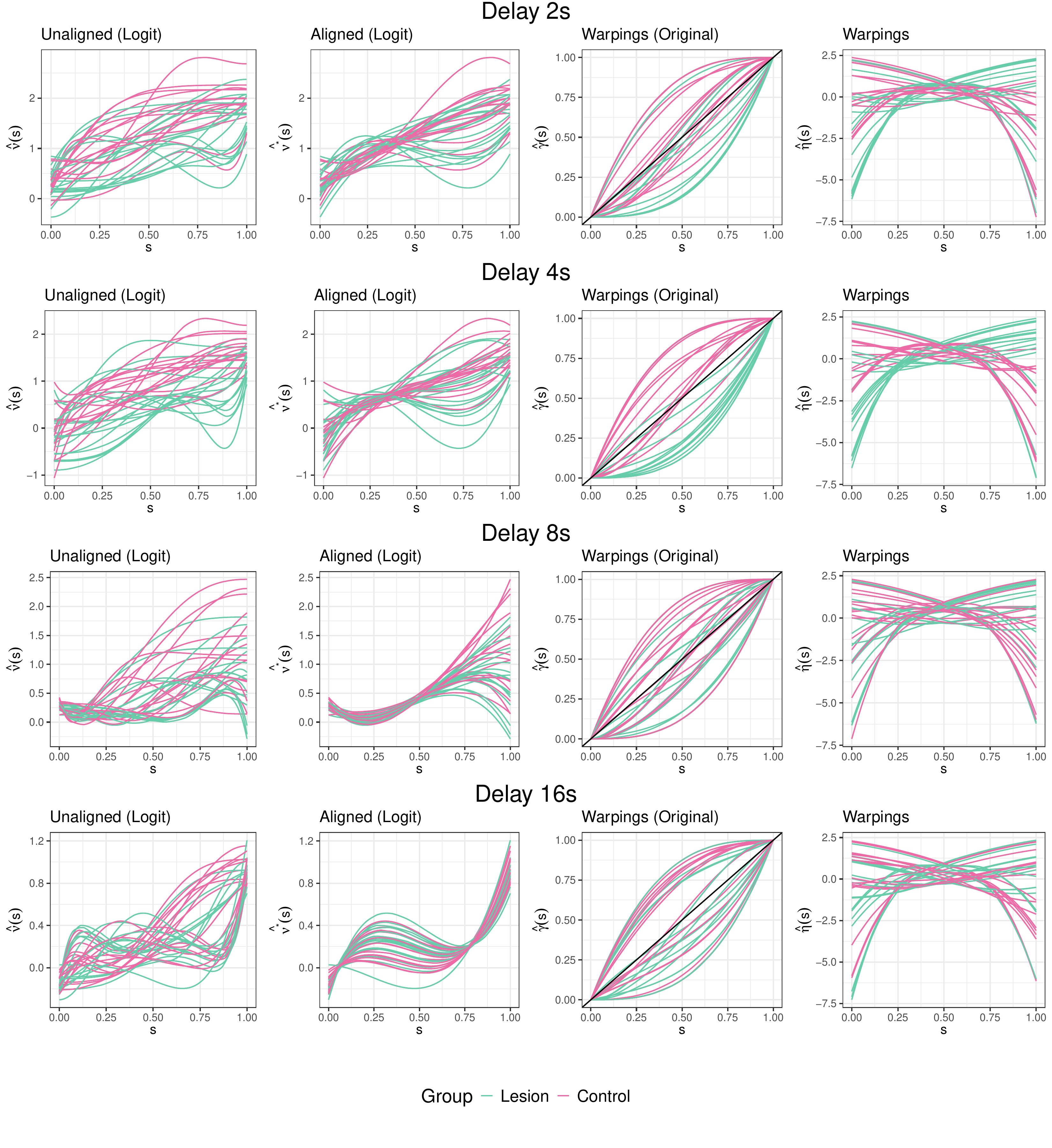}
  \caption{Functional samples when delays are introduced. We show the logit-transformed learning curves (first panel), their aligned version (second panel), the warping functions of the alignment (third panel), and their transformation (fourth panel). In all samples, lesion mice are colored in teal, and control mice in pink.}
  \label{fig:delays_sample}
\end{figure}

\subsection{Test statistic of the permutation tests}
\begin{figure}[htb!]
  \centering
\includegraphics[width=0.8\textwidth]{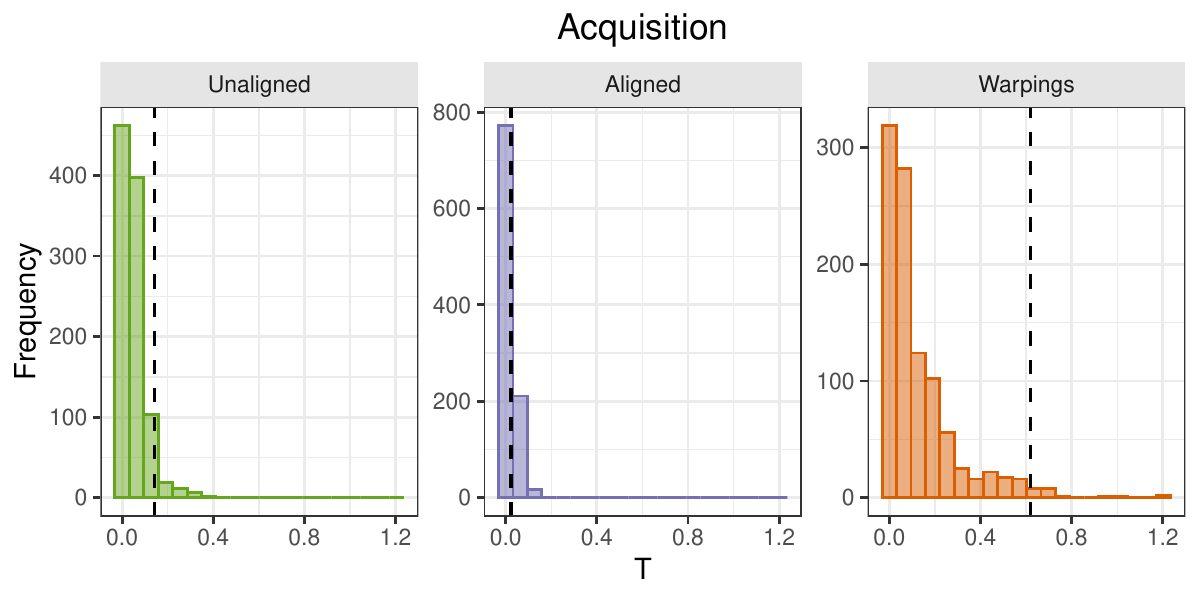}
  \caption{Histograms of test statistic $T$ for 1000 permutations, divided by types of curves: original (top), aligned (middle) and transformed warping functions (bottom). The vertical line indicates the observed statistic for the original datasets.}
  \label{fig:T_perm_learning}
\end{figure}

\begin{figure}[htb!]
  \centering
\includegraphics[width=\textwidth]{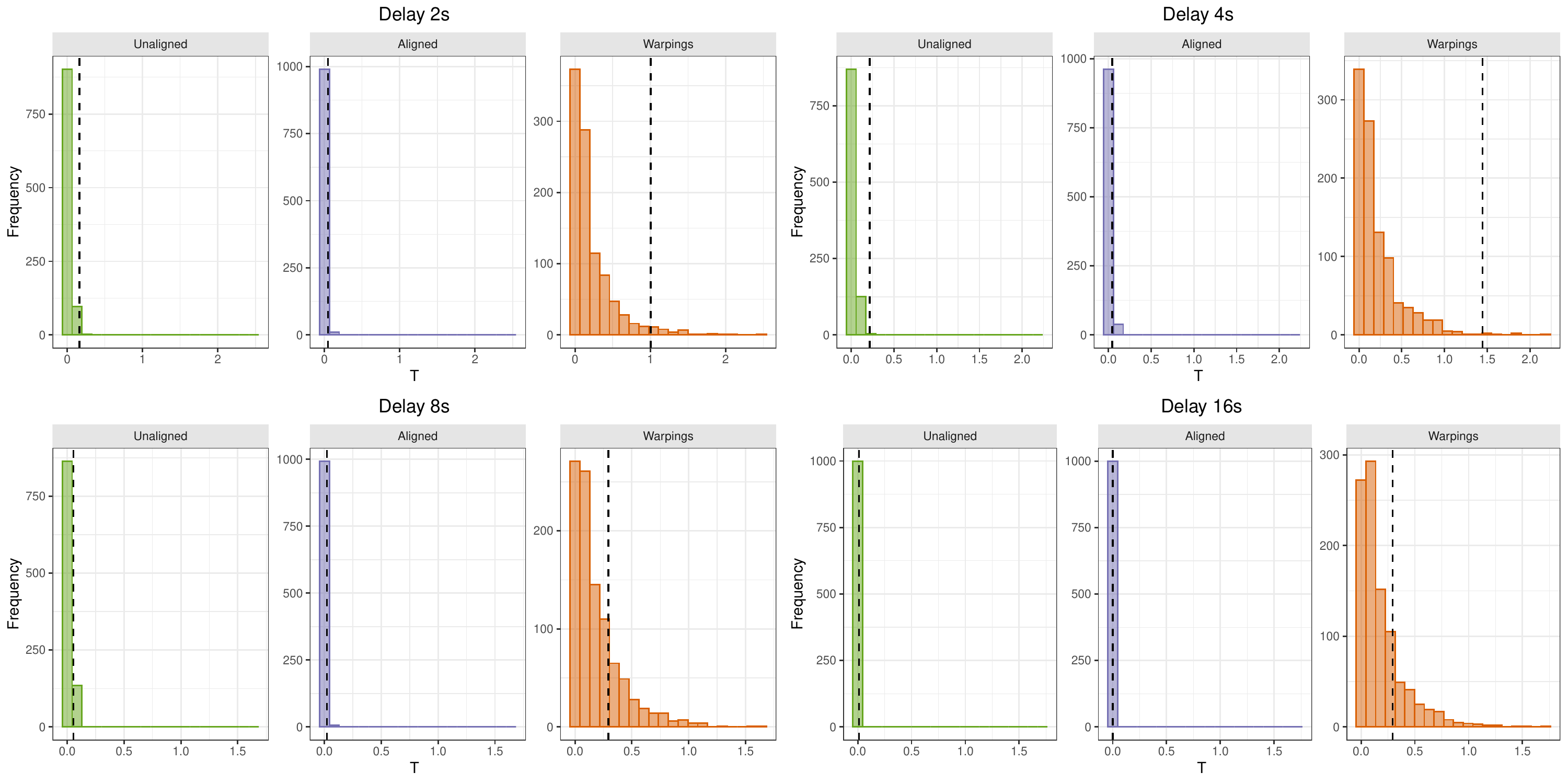}
   \caption{Histograms of test statistic $T$ for 1000 permutations corresponding to delay type, divided by types of curves: original (top), aligned (middle) and transformed warping functions (bottom). The vertical line indicates the observed statistic for the original datasets. }
  \label{fig:T_perm_memory}
\end{figure}

In Figures \ref{fig:T_perm_learning} and \ref{fig:T_perm_memory} we show the distributions of the permutation tests statics for the learning and memory tasks, respectively. They correspond to the $p$-values in Table \ref{tab:perm_tests}.

\end{appendices}

\end{document}